\documentclass[twocolumn,showpacs,aps,prl]{revtex4-1}

\usepackage{amsmath,amsfonts,amssymb}
\usepackage{wrapfig}
\usepackage{graphicx}
\usepackage{bbm}
\usepackage{color}

\begin{document}

\title{Genuine High-Order Einstein-Podolsky-Rosen Steering}\date{\today}
\author{Che-Ming Li$^{1}$}
\email{cmli@mail.ncku.edu.tw}
\author{Kai Chen$^{2,3}$}
\email{kaichen@ustc.edu.cn}
\author{Yueh-Nan Chen$^{4}$}
 \author{Qiang Zhang$^{2,3}$}
 \author{Yu-Ao Chen$^{2,3}$}
\author{Jian-Wei Pan$^{2,3}$}
\email{pan@ustc.edu.cn}

\affiliation{$^1$Department of Engineering Science, National Cheng Kung University, Tainan 701, Taiwan}
\affiliation{$^2$Hefei National Laboratory for Physical Sciences at Microscale and Department of Modern Physics,
University of Science and Technology of China, Hefei, Anhui 230026, China}
\affiliation{$^3$Shanghai Branch, CAS Center for Excellence and Synergetic Innovation Center in Quantum Information and Quantum Physics, University of Science and Technology of China, Shanghai 201315, China}
\affiliation{$^4$Department of Physics and National Center for Theoretical Sciences, National Cheng-Kung University, Tainan 701, Taiwan}


\begin{abstract}
Einstein-Podolsky-Rosen (EPR) steering demonstrates that two parties share entanglement even if the measurement devices of one party are untrusted. Here, going beyond this bipartite concept, we develop a novel formalism to explore a large class of EPR steering from generic multipartite quantum systems of arbitrarily high dimensionality and degrees of freedom, such as graph states and hyperentangled systems. All of these quantum characteristics of genuine high-order EPR steering can be efficiently certified with few measurement settings in experiments. We faithfully demonstrate for the first time such generality by experimentally showing genuine four-partite EPR steering and applications to universal one-way quantum computing. Our formalism provides a new insight into the intermediate type of genuine multipartite Bell nonlocality and potential applications to quantum information tasks and experiments in the presence of untrusted measurement devices.
\end{abstract}

\maketitle

Einstein-Podolsky-Rosen (EPR) steering was originally introduced by Schr\"odinger in 1935 \cite{Shrodinger35} to describe the EPR paradox \cite{EPR}. Recently, it has been formulated by Wiseman \textit{et al.} \cite{Wiseman07} to show a strict hierarchy among Bell nonlocality, steering, and entanglement \cite{Cavalcanti09,Saunders10} and stimulated new applications to quantum communication \cite{Branciard12}. Several experimental demonstrations of EPR steering have been reported \cite{Saunders10,Smith12,Wittmann12}. The steering effect reveals that different ensembles of quantum states can be remotely prepared by measuring one particle of an entangled pair. We go a step further and consider the following question: how to experimentally observe genuine multipartite EPR steering? For instance, given an experimental output state $\rho_{\text{expt}}$ which is created according a target four-qubit cluster state of the form \cite{Walther05,Vallone07}
\begin{eqnarray}
\left|G_{4}\right\rangle&=&\frac{1}{2}(\left|+\right\rangle_{1}\left|0\right\rangle_{2}\left|+\right\rangle_{3}\left|0\right\rangle_{4}+\left|-\right\rangle_{1}\left|1\right\rangle_{2}\left|+\right\rangle_{3}\left|1\right\rangle_{4}\nonumber\\
&&\ \ \  +\left|+\right\rangle_{1}\left|0\right\rangle_{2}\left|-\right\rangle_{3}\left|1\right\rangle_{4}+\left|-\right\rangle_{1}\left|1\right\rangle_{2}\left|-\right\rangle_{3}\left|0\right\rangle_{4}),\nonumber
\end{eqnarray}
where $\left|\pm\right\rangle_{k}=(\left|0\right\rangle_{k}\pm\left|1\right\rangle_{k})/\sqrt{2}$, how do we describe the effect of genuine multipartite EPR steering and then detect such steerability of $\rho_{\text{expt}}$ in the laboratory?

Inspired by the task-oriented formulation of bipartite steering \cite{Wiseman07}, genuine multipartite EPR steering can be defined from an operational interpretation as the distribution of multipartite entanglement by \emph{uncharacterized} (or untrusted) parties. Let us consider a system composed of $N$ parties and a source creating $N$ particles. Each party of the system can receive a particle from the source whenever an $N$-particle state is created. We divide the system into two groups, say $A_{s}$ and $B_{s}$, and assume that $A_{s}$ is responsible for sending particles from such a source to every party. Each time, after receiving particles, they measure their respective parts and communicate classically. Since $B_{s}$ does not trust $A_{s}$, $A_{s}$'s task is to convince $B_{s}$ that the state shared between them is entangled. $A_{s}$ will succeed in this task if and only if $A_{s}$ can prepare different ensembles of quantum states for $B_{s}$ by steering $B_{s}$'s state. Here we say an $N$-particle state generated from the source to possess genuine $N$-partite EPR steerability if by which $A_{s}$ succeed in the task for \emph{all} possible bipartitions $A_{s}$ and $B_{s}$ of the $N$-particle system. This interpretation is consistent with the definition recently introduced by He and Reid \cite{He13}. In a wider scope, Schr\"odinger's original concept can even be applied to quantum systems with many degrees of freedom (DOFs), e.g., hyperentangled systems, and, as will be shown presently, extended as \emph{genuine} multi-DOF EPR steering. In this Letter we call these two sorts of quantum characteristics genuine high-order EPR steering.

\begin{figure*}[t]
\includegraphics[width=17.7 cm,angle=0]{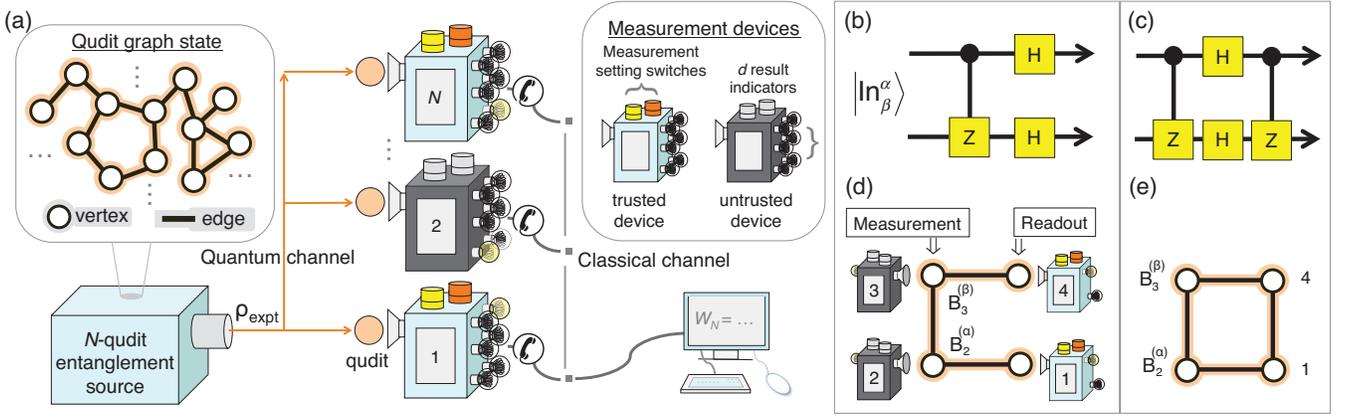}
\caption{(color online). Genuine multipartite EPR steerability and applications. Generic entanglement-based quantum protocols rely on both characterized measurement devices and genuine multipartite entanglement, such as one-way quantum computation \cite{Raussendorf01b,Raussendorf03} and multiparty quantum communications \cite{Hillery99}. Genuine multipartite EPR steerability enables them to perform quantum applications in the presence of untrusted measurement apparatus (a). Here, we develop quantum steering witnesses to ensure that an experimental state $\rho_{\text{expt}}$ of an $N$-partite quantum $d$-dimensional system ($N$ qudits) has such ability, for example, an experimental graph state \cite{Raussendorf01a,Hein04,Zhou03,SI} used for one-way computing. To implement gate operations such as circuits, composed of one-qubit ($d=2$) gate $\hat{H}$ and two-qubit controlled-$Z$ (CZ) gates (b) and (c) for input states $\left|\text{In}^{\alpha}_{\beta}\right\rangle$, one needs to prepare chain-type (d), i.e., $\left|G_{4}\right\rangle$, and box-type (e) cluster states, respectively.
By performing measurements $B_{2}^{(\alpha)}$ and $B_{3}^{(\beta)}$ on qubits $2$ and $3$, respectively, the rest of the qubits $1,4$ together with the outcomes of measurements on qubits $2$, $3$ would provide a readout of the gate operations. See the Supplemental Material for detailed discussions \cite{SI}. In addition to releasing the assumptions about the measurement devices, genuine multipartite steerability promises high-quality one-way computation (see Fig.~\ref{RES}).}
\label{GEPR}
\end{figure*}  

The concept of verifying genuine high-order EPR steering leads us naturally to consider quantum scenarios based on genuine high-order entanglement \cite{Ma11,Giovannetti11,Hillery99,Raussendorf01b,Raussendorf03,Raussendorf07} in which $B_{s}$'s measurement apparatus are trusted, while $A_{s}$'s are not; see Fig.~\ref{GEPR}. Demonstrations of genuine high-order EPR steerability guarantee faithful implementations of the quantum scenarios in the presence of uncharacterized parties. So far, while verifying genuine tripartite steering becomes possible \cite{He13}, the fundamental problem such as the verification considered above and the cases for arbitrary large $N$ remains open.

Here we develop new \emph{quantum witnesses} to observe a large class of genuine high-order EPR steering independent of the particle or DOF number. Let us start with a demonstration of the first experimental genuine four-partite EPR steerability for states close to the cluster state $\left|G_{4}\right\rangle$. In our scenario, we assume that two possible measurements can be performed on each particle ($m_{k}=1,2$ for the $k$th particle) and that each local measurement has two possible outcomes, $v_{k}^{(m_{k})}\in\{0,1\}$. We take the measurements for each party who implements quantum measurements to observables with the nondegenerate eigenvectors $\{\left|0\right\rangle_{k,1}=\left|0\right\rangle_{k},\left|1\right\rangle_{k,1}=\left|1\right\rangle_{k}\}$ for $m_{k}=1$ and $\{\left|0\right\rangle_{k,2}=\left|+\right\rangle_{k},\left|1\right\rangle_{k,2}=\left|-\right\rangle_{k}\}$ for $m_{k}=2$.

The genuine four-partite EPR steerability of the ideal cluster state $\left|G_{4}\right\rangle$ is revealed by the following relations of measurement results: $v_{1}^{(1)}+v_{2}^{(2)}+v_{3}^{(1)}\doteq 0$ and $v_{3}^{(1)}+v_{4}^{(2)}\doteq 0$, and the relation: $v_{2}^{(1)}+v_{3}^{(2)}+v_{4}^{(1)}\doteq 0$ and $v_{1}^{(2)}+v_{2}^{(1)}\doteq 0$, where $\doteq$ denotes equality modulo $2$. When $A_{s}$ and $B_{s}$ share a state $\left|G_{4}\right\rangle$, $A_{s}$ can steer the states of $B_{s}$'s particles by measuring the particles held as described by the above relations whatever the partition $A_{s}$ and $B_{s}$ is considered. Thus we construct the kernel of quantum steering witness in the form
\begin{eqnarray}
W_{4} &\equiv& P(v_{1}^{(1)}+v_{2}^{(2)}+v_{3}^{(1)}\doteq 0,v_{3}^{(1)}+v_{4}^{(2)}\doteq 0)\nonumber\\
&& +P(v_{2}^{(1)}+v_{3}^{(2)}+v_{4}^{(1)}\doteq 0,v_{1}^{(2)}+v_{2}^{(1)}\doteq 0),\nonumber
\end{eqnarray}
where $P(\cdot)$ denotes the joint probability of obtaining experimental outcomes satisfying the designed conditions. If $B_{s}$ has a preexisting state known to $A_{s}$, rather than part of a genuine multipartite entanglement shared with $A_{s}$, the maximum value of $W_{4}$ is
\begin{eqnarray}
W_{4C} \equiv \max_{\mathbf{A}_{s},\{\text{v}_{a}\}_{C},\{\text{v}_{b}\}_{QM}}W_{4}=1+\frac{1}{\sqrt{2}}\sim 1.7071\nonumber,
\end{eqnarray}
where $\mathbf{A}_{s}$ denotes the index set of $A_{s}$ for all possible partitions and $\{\text{v}_{a}\}_{C}$ indicates that the outcome set for $A_{s}$ is derived from such a preexisting-state scenario. The outcome set $\{\text{v}_{b}\}_{QM}$ of $B_{s}$ is obtained by performing quantum measurements on the preexisting quantum states. Hence we posit that if an experimental state $\rho_{\text{expt}}$ shows that
\begin{equation}
W_{4}(\rho_{\text{expt}})>1+\frac{1}{\sqrt{2}},\label{w4}
\end{equation}
then $\rho_{\text{expt}}$ can exhibit genuine four-partite EPR steerability close to $\left|G_{4}\right\rangle$. This certification rules out all the possibilities of results mimicked by tripartite steerability, including all possible mixtures of them. States certified by this witness will enable the quantum protocols to be implemented even when uncharacterized measurement apparatus are unavoidably used (Fig.~\ref{GEPR}). It is also worth noting that the steering witness~(\ref{w4}) is efficient. Only the \emph{minimum} two local measurement settings are sufficient to show the steerability.

 \begin{figure}[t]
\includegraphics[width=8.7 cm,angle=0]{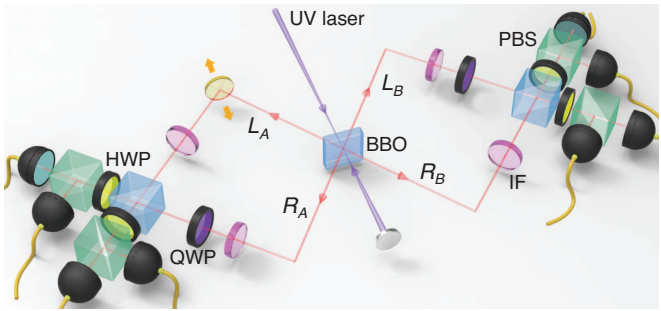}
\caption{(color online). Experimental setup. A pulse (5 ps) of UV light with a central wavelength of 355 nm and an average power of 200 mW at repetition rate of 80 MHz double passes a two-crystal structured BBO to produce polarization entangled photon pairs either in the forward direction or in the backward direction. To create desired entangled pairs in mode $R_{A},R_{B}$ and in $L_{A},L_{B}$, two quarter wave plates (QWPs) are properly tilted along their optic axis. Half wave plates (HWPs), polarizing beam splitters (PBSs), and eight single-photon detectors are used as polarization analyzers for the output states. Here $3$-nm bandpass filters (IFs) with central wavelength $710$ nm are placed in front of them.}
\label{EXP}
\end{figure}

\begin{figure*}[t]
\includegraphics[width=17.7 cm,angle=0]{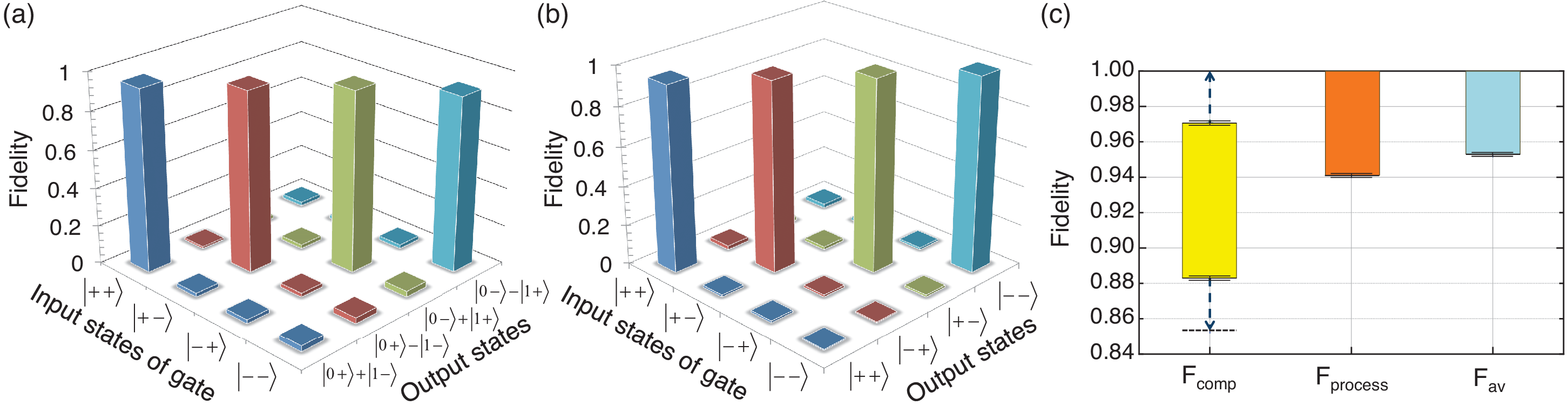}
\caption{(color online). Steerability and performance of quantum gates. Two kinds of gate operations are experimentally realized for the target quantum gates [Figs.~\ref{GEPR}(b),\ref{GEPR}(c)]. We measure the fidelities of the output states by inputting four orthogonal states into the experimental gates, $\left|mn\right\rangle$ for $m,n=+,-$ \cite{FIG3}: (a) a mean fidelity $\sim0.935\pm0.004$ for the gate operation, Fig.~\ref{GEPR}(b), and (b) an average fidelity $\sim 0.960\pm 0.004$ for the target gate, Fig.~\ref{GEPR}(c). (c) From the estimate of the average computation fidelity $F_{\text{\text{comp}}}$ (\ref{fidelity}), we obtain the lower bounds of the quantum process fidelity $F_{\text{process}}$ ($\sim 0.9415\pm 0.0025$) and the average state fidelity $F_{\text{av}}$ ($\sim 0.9532\pm0.0020$), which indicate good qualities of our experimental gates, regardless of the input states \cite{FIG3}. Since both created gates are based on the same source, their estimations of the three different fidelities are identical \cite{SI}. The gate performance reveals genuine four-partite steerability by the steering witness $F_{\text{comp}}>W_{4C}/2\sim 0.8536$. The double-arrow dashed line shows the region where the source states are truly four-partite steerable.}
\label{RES}
\end{figure*}

To experimentally observe the genuine-four partite EPR steerability, we utilize the technique developed in the previous experiment \cite{Kai07} to generate a source of two-photon four-qubit cluster states entangled both in polarization and spatial modes. As illustrated in Fig.~\ref{EXP}, an ultraviolet (UV) pulse passes twice through two contiguous type-I-$\beta$ barium borate (BBO) to produce polarization entangled photon pairs in the forward (spatial modes $R_{A,B}$) and the backward ($L_{A,B}$) directions. Through temporal overlaps of modes $R_{A}$ and $L_{A}$ and of modes $R_{B}$ and $L_{B}$, one can create a four-qubit state
\begin{eqnarray}
\left|G'_{4}\right\rangle&=&\frac{1}{2}\big[(\left|H\right\rangle_{A}\left|H\right\rangle_{B}+\left|V\right\rangle_{A}\left|V\right\rangle_{B})\left|R\right\rangle_{A}\left|R\right\rangle_{B}\nonumber\\
&&\ \ \  +(\left|H\right\rangle_{A}\left|H\right\rangle_{B}-\left|V\right\rangle_{A}\left|V\right\rangle_{B})\left|L\right\rangle_{A}\left|L\right\rangle_{B}\big],
\end{eqnarray}
entangled both in spatial modes and horizontal ($H$) and vertical ($V$) polarizations. By encoding logical qubits as $\left|H(V)\right\rangle_{A(B)}\equiv\left|0(1)\right\rangle_{1(2)}$ and $\left|R(L)\right\rangle_{A(B)}\equiv\left|0(1)\right\rangle_{3(4)}$, $\left|G'_{4}\right\rangle$ is equivalent to the cluster state $\left|G_{4}\right\rangle$ up to a transformation $\hat{H}_{1}\otimes\hat{H}_{4}$ where $\hat{H}_{k}\left|0(1)\right\rangle_{k}=\left|+(-)\right\rangle_{k}$. The witness kernel for this target state has a corresponding change in measurement settings by $W'_{4} \equiv P(v_{1}^{(2)}+v_{2}^{(2)}+v_{3}^{(1)}\doteq 0,v_{3}^{(1)}+v_{4}^{(1)}\doteq 0)+P(v_{2}^{(1)}+v_{3}^{(2)}+v_{4}^{(2)}\doteq 0,v_{1}^{(1)}+v_{2}^{(1)}\doteq 0)$.

In the experiment, we obtain a high generation rate of cluster state about $1.2\times 10^{4}$ per sec with $200$ mW UV pump. We measure the two joint probabilities in the witness kernel with the designed measurement settings \cite{LMS} and have $P(v_{1}^{(2)}+v_{2}^{(2)}+v_{3}^{(1)}\doteq 0,v_{3}^{(1)}+v_{4}^{(1)}\doteq 0) = 0.9490\pm 0.0022$ and $P(v_{2}^{(1)}+v_{3}^{(2)}+v_{4}^{(2)}\doteq 0,v_{1}^{(1)}+v_{2}^{(1)}\doteq 0)= 0.9339\pm 0.0027$. Then we observe genuine four-partite steerability verified by
\begin{eqnarray}
W'_{4}(\rho_{\text{expt}})&=&1.8829\pm 0.0049.\label{wexp}
\end{eqnarray}
This result is clearly larger than the maximum value the preexisting-state scenario can achieve, $W_{4C}$, by $\sim36$ standard deviations. 

The measured witness kernel $W_{4}(\rho_{\text{expt}})$ can also reveal the information about the state fidelity, $F_{S}(\rho_{\text{expt}})= \text{Tr}[\left|G_{4}\right\rangle\left\langle G_{4}\right|\rho_{\text{expt}}]$. To derive such connection, let us first represent the witness (\ref{w4}) in the operator form, $\hat{W}_{G_{4}}\equiv W_{4C}\hat{I}-\hat{\mathcal{W}}_{4}$, where  $\hat{I}$ denotes the identity operator and $\hat{\mathcal{W}}_{4}$ corresponds to $W_{4}$ such that $W_{4}(\rho_{\text{expt}})=\text{Tr}[\hat{\mathcal{W}}_{4}\rho_{\text{expt}}]$.  If $\langle\hat{W}_{G_{4}}\rangle=\text{Tr}[\hat{W}_{G_{4}}\rho_{\text{expt}}]<0$, then $\rho_{\text{expt}}$ is genuinely four-partite steerable. One can use $\hat{W}_{G_{4}}$ to derive a steering witness operator of the form $\hat{W}'_{G_{4}}\equiv W_{4C}/2\hat{I}-\left|G_{4}\right\rangle\left\langle G_{4}\right|$ \cite{SI}, which provides a steering witness $F_{S}(\rho_{\text{expt}})>W_{4C}/2$ $(\sim 0.8536)$. They satisfy the relation $\hat{W}'_{G_{4}}-\gamma\hat{W}_{G_{4}}\geq 0$, where $\gamma$ is some positive constant, which means that when a state is detected by $\hat{W}'_{G_{4}}$, it is certified by $\hat{W}_{G_{4}}$ as well. Such a relation can be manipulated to give $\left|G_{4}\right\rangle\left\langle G_{4}\right |\leq\hat{\mathcal{W}}_{4}/2$, i.e., to provide the upper bound of the state fidelity, $F_{S}(\rho_{\text{expt}})\leq W_{4}(\rho_{\text{expt}})/2$. Similarily, we use $\hat{W}'_{G_{4}}$ to construct another steering witness operator composed of $\hat{\mathcal{W}}_{4}$ \cite{SI} and derive the lower bound of $F_{S}(\rho_{\text{expt}})$: $\hat{\mathcal{W}}_{4}-\hat{I}\leq \left|G_{4}\right\rangle\left\langle G_{4}\right |$. Hence, $ F_{S}(\rho_{\text{expt}})$ is estimated as
\begin{equation}
0.8829\pm 0.0049\leq  F_{S}(\rho_{\text{expt}})\leq 0.9415\pm 0.0025.\label{StateFidelity}
\end{equation}

Such characteristic of genuine four-partite steering serves as a source for implementing faithful one-way quantum computing. We have realized the quantum gates illustrated in Fig.~\ref{GEPR} \cite{SI} and objectively evaluated their performance and connection with steerability. We use $W'_{4}(\rho_{\text{expt}})$ (\ref{wexp}) to estimate three different fidelities \cite{FIG3}: their average computation fidelity ($F_{\text{\text{comp}}}$) \cite{Hofmann05}, the quantum process fidelity ($F_{\text{process}}$) \cite{Gilchrist05}, and the average state fidelity ($F_{\text{av}}$) \cite{Gilchrist05}; see Fig.~\ref{RES}. To connect steerability with the gate operation shown in Fig~\ref{GEPR}(b), we first construct a steering witness operator of the form $\hat{W}_{\text{CZ}}\equiv W_{4C}\hat{I}-\hat{\mathcal{W}}_{\text{CZ}}$ \cite{SI}, where the operator $\hat{\mathcal{W}}_{\text{CZ}}$ specifies the relation between ideal input and output states of the quantum circuit \cite{CZ}. This means the steerability can be verified by performing one-way computation to check $\langle\hat{\mathcal{W}}_{\text{CZ}}\rangle> W_{4C}$. Following the same method as that used to estimate the lower and upper bounds of $F_{S}(\rho_{\text{expt}})$ \cite{SI}, we get $W_{4}(\rho_{\text{expt}})-1\leq\langle\hat{\mathcal{W}}_{\text{CZ}}\rangle\leq W_{4}(\rho_{\text{expt}})/2+1$. Hence, with the relation $F_{\text{\text{comp}}}=\langle\hat{\mathcal{W}}_{\text{CZ}}\rangle/2$ \cite{CZ} and the estimation of $\langle\hat{\mathcal{W}}_{\text{CZ}}\rangle$, we arrive at the steering witness $F_{\text{\text{comp}}}> W_{4C}/2$ and the fidelity estimation 
\begin{equation}
W_{4}(\rho_{\text{expt}})-1\leq F_{\text{comp}}\leq \frac{1}{4}W_{4}(\rho_{\text{expt}})+\frac{1}{2}.\label{fidelity}
\end{equation}
$F_{\text{process}}$ and $F_{\text{av}}$ are further determined by  $F_{\text{\text{comp}}}$ \cite{FIG3}.

The concept and method of the criterion (\ref{w4}) can be directly applied to quantum states with complex structures. One of the extensions is to certify steerability of a general $d$-dimensional $N$-partite and $q$-colorable graph state  $\left|G\right\rangle$ \cite{Raussendorf01a,Hein04,Zhou03} [Fig.~\ref{GEPR}(a)]. The kernel of the steering witness is of the form
\begin{equation}
W_{N}(q,d)\equiv\sum_{m=1}^{q}P(v^{(2)}_{j}+\sum_{i\in \epsilon(j)}v^{(1)}_{i}\doteq 0|\forall j \in Y_{m}),\nonumber
\end{equation}
for $v^{(1)}_{i},v^{(2)}_{j}\in\{0,1,...,d-1\}$ measured in two complementary bases \cite{SI}, where $\doteq$ denotes equality modulo $d$ and $\epsilon(j)$ represents the set of vertices that form edges with the vertex $j$ in the color set $Y_{m}$. The quantum steering witness is described by
\begin{equation}
W_{N}(q,d|\rho_{\text{expt}})>\frac{1}{2}(q+\sqrt{\frac{2(q^2-2q+2)+\gamma_{q}}{d}}),\label{wn}
\end{equation}
where $\gamma_{2}=0$ and $\gamma_{q}=[2(q-3)+1]+\gamma_{q-1}$ for $q\geq 3$ \cite{SI}. If $\rho_{\text{expt}}$ is detected by Eq.~(\ref{wn}), then $\rho_{\text{expt}}$ possesses genuine $N$-partite EPR steerability close to $\left|G\right\rangle$. This witness has the same features as the witness (\ref{w4}) \cite{WitnessFeatures} and possesses high robustness against noise \cite{SI}. With only $q$ local measurement settings, $W_{N}(q,d)$ can be efficiently realized regardless of the number of qudits. We remark that, for states that do not belong to the above state types, for example, the $W$ states \cite{Dur00}, useful steering witnesses still could be derived from one's knowledge to this target state \cite{SI}.  For more extensions, how to observe EPR steering in \emph{all} DOFs \cite{Kwiat97,Barreiro05} under consideration is shown in the Supplemental Material. The concrete experimental illustrations and applications of such genuine high-order EPR steering are also detailed therein \cite{SI}.

In conclusion, we have developed a novel formalism to explore genuine high-order EPR steering and experimentally demonstrated such generality and applications with photonic cluster states. Being capable of revealing genuine high-order EPR steering pushes beyond the capability of bipartite steering and promotes potential applications and experiments. One can probe more sorts of steerability, for example, in resonating valence-bond states, which would allow an analogue quantum simulator \cite{Ma11} to run without fully characterized quantum measurements. Similarly, other quantum strategies based on both characterized measurements and genuine multipartite entanglement, like quantum metrology \cite{Giovannetti11}, can benefit from it as well. Moreover, since genuine multipartite EPR steerability cannot be mimicked by parts of the whole system, such ability together with steering witness could facilitate multipartite secret sharing \cite{Hillery99} in a generic one-sided device-independent mode \cite{Branciard12,He13}. It is interesting to compare our method with the recently developed single-system steering for quantum information processing \cite{Li14} and to investigate further from the all-versus-nothing (AVN) point of view. Subtle AVN proof for steering and their experimental demonstration are given for special classes of two qubits \cite{ChenJL13,GuoAVN14}.

We thank Q.-Y. He, J.-L. Chen, and Y.-C. Liang for discussions. This work has been supported by the Chinese Academy of Science, the National Fundamental Research Program, and the National Natural Science Foundation of China. C.-M. L. acknowledges partial support from the Ministry of Science and Technology, Taiwan, under Grant No. MOST 101-2112-M-006-016-MY3. Y.-N. C. acknowledges partial support from the National Center for Theoretical Sciences and the Ministry of Science and Technology, Taiwan, under Grant No. MOST 103-2112-M-006-017-MY4.

\textit{Note added.---} Recently, we became aware of two experimental demonstrations of tripartite steering \cite{Armstrong14,Cavalcanti14}.


\begin{references}
\bibitem{LMS} We use three different setups to implement spatial mode analysis. When performing the measurement $m_{k}=2$ for $k=3$ ($4$), a beam splitter is set at the intersection of modes $R_{A(B)}$ and $L_{A(B)}$. Furthermore, combined with $\alpha$-phase shifter, measurement $B_{k}^{(\alpha)}$ can be realized for one-way quantum computing [Figs.~\ref{GEPR}(d),(e) and \cite{SI}]. Whereas for the measurement $m_{k}=1$, they are left empty. If a PBS is placed at the mode intersection, one can realize the measurements $m_{1,2}=1$ and $m_{3,4}=1$ simultaneously.
\end{references}
\end{document}


\title{Genuine High-Order Einstein-Podolsky-Rosen Steering: Supplementary Material}\date{\today}
\author{Che-Ming Li$^{1}$}
\email{cmli@mail.ncku.edu.tw}
\author{Kai Chen$^{2,3}$}
\email{kaichen@ustc.edu.cn}
\author{Yueh-Nan Chen$^{4}$}
 \author{Qiang Zhang$^{2,3}$}
 \author{Yu-Ao Chen$^{2,3}$}
\author{Jian-Wei Pan$^{2,3}$}
\email{pan@ustc.edu.cn}

\affiliation{$^1$Department of Engineering Science, National Cheng Kung University, Tainan 701, Taiwan}
\affiliation{$^2$Hefei National Laboratory for Physical Sciences at Microscale and Department of Modern Physics,
University of Science and Technology of China, Hefei, Anhui 230026, China}
\affiliation{$^3$Shanghai Branch, CAS Center for Excellence and Synergetic Innovation Center in Quantum Information and Quantum Physics, University of Science and Technology of China, Shanghai 201315, China}
\affiliation{$^4$Department of Physics and National Center for Theoretical Sciences, National Cheng-Kung University, Tainan 701, Taiwan}

\begin{abstract}
In this supplementary information, we provide the required material to support our theoretical and experimental results presented in the main text. Novel applications of our theoretical scheme to the existing experiments are supplemented as well. Firstly, we describe in detail the basic principle of our experimental quantum gates. Then, we provide a complete proof of the introduced quantum steering witness for genuine multipartite Einstein-Podolsky-Rosen (EPR) steerability of states close to a $d$-dimensional $N$-partite and $q$-colorable graph state. The robustness of the steering witness is discussed. A concrete illustration of genuine $N$-degree-of-freedom EPR steering and certification of such genuine high-order EPR steering are introduced in detail. In addition, we prove that the state fidelity of a generated state with respect to the target state $F_{S}(\rho_{\text{expt}})$ can be estimated from the measured steering witness kernel, and furthermore, that $F_{S}(\rho_{\text{expt}})$ can serve as an indicator showing genuine multipartite steerability. We also show that the steerability can be detected by performing one-way quantum computing and that how we use the measured witness kernel to estimate the average computation fidelity ($F_{\text{\text{comp}}}$), the quantum process fidelity ($F_{\text{process}}$), and the average state fidelity ($F_{\text{av}}$). Finally, a method of constructing quantum steering witnesses based on full state knowledge is introduced in the end. 
\end{abstract}

\pacs{03.65.Ud, 03.67.Lx, 42.50.Dv}
\maketitle

\textit{Experimental quantum gates.---}For a given cluster state, a quantum computing is defined by consecutive single-qubit measurements in basis $B_{k}^{(\alpha)}=\{\left|\alpha_{+}\right\rangle_{k},\left|\alpha_{-}\right\rangle_{k}\}$ and their results of measurements, where $\left|\alpha_{\pm}\right\rangle_{k}=(\left|0\right\rangle_{k}\pm e^{i\alpha}\left|1\right\rangle_{k})/\sqrt{2}$ for any real $\alpha$ \cite{Raussendorf01b,Raussendorf03}. A measurement outcome of $\left|\alpha_{+}\right\rangle_{k}$ denotes $s_{k}=0$ while $\left|\alpha_{-}\right\rangle_{k}$ signifies $s_{k}=1$. This measurement basis determines a gate operation $R_{z}^{(\alpha)}=\exp(-i\alpha Z/2)$ where $Z\equiv \left|0\right\rangle\left\langle 0\right|-\left|1\right\rangle\left\langle 1\right|$ \cite{Nielsen00}, followed by a transformation $\hat{H}$. A chain-type graph (horseshoe cluster) state $\left|G_{4}\right\rangle$ [Fig.~1(d) in the main text] can be used to realize a two-qubit controlled-$Z$ gate ($U_{\text{CZ}}$) [Fig.~1(b)]: $U_{\text{CZ}}\left|j\right\rangle\left|k\right\rangle=(-1)^{jk}\left|j\right\rangle\left|k\right\rangle$, for $j,k=0,1$. When measuring along bases $B_{2}^{(\alpha)}$ and $B_{3}^{(\beta)}$, the state of the qubits $1$ and $4$ would be
\begin{equation}
(X^{s_{2}}\otimes X^{s_{3}})(\hat{H}\otimes\hat{H})U_{\text{CZ}}R_{z}^{(-\alpha)}\otimes R_{z}^{(-\beta)}\left|+\right\rangle\left|+\right\rangle.\nonumber
\end{equation}
Let us focus on measurements with the outcome $s_{2}=0$ and $s_{3}=0$. It is easy to find that, for such a case, the post-measurement state of the 2nd and the 3rd qubits, $\left|\alpha_{+}\right\rangle_{2}\otimes\left|\beta_{+}\right\rangle_{3}$, determines the input state and the connection between input and output states of the quantum gate $(\hat{H}\otimes\hat{H})U_{\text{CZ}}$ by
\begin{equation}
\left|\text{Out}^{\alpha}_{\beta}\right\rangle=(\hat{H}\otimes\hat{H})U_{\text{CZ}}\left|\text{In}^{\alpha}_{\beta}\right\rangle,\tag{S1}\label{output}
\end{equation}
where
\begin{equation}
\left|\text{In}^{\alpha}_{\beta}\right\rangle=R_{z}^{(-\alpha)}\otimes R_{z}^{(-\beta)}\left|+\right\rangle\left|+\right\rangle=\left|-\alpha_{+}\right\rangle\left|-\beta_{+}\right\rangle.\tag{S2}\label{input}
\end{equation}
Then, one can input different states $\left|\text{In}^{\alpha}_{\beta}\right\rangle$ into the gate to see the effect of the gate operation $(\hat{H}\otimes\hat{H})U_{\text{CZ}}$ by analyzing the output states $\left|\text{Out}^{\alpha}_{\beta}\right\rangle$. For examples, one can set the angles $(\alpha,\beta)$ as $(0,0)$, $(0,\pi)$, $(\pi,0)$ and $(\pi,\pi)$ to prepare an orthonormal set of the input states $\left|+\right\rangle\left|+\right\rangle$, $\left|+\right\rangle\left|-\right\rangle$, $\left|-\right\rangle\left|+\right\rangle$, and $\left|-\right\rangle\left|-\right\rangle$, respectively. Here, for $(\alpha, \beta)\in\{0,\pi\}$, the output states are entangled [Fig.~3(a)]. One can also design another orthonormal set of input states which is complementary to the above set of input states \cite{Hofmann05}. As $(\alpha,\beta)$ are chosen as $(-\pi/2,-\pi/2)$, $(-\pi/2,\pi)$, $(\pi/2,-\pi/2)$ and $(\pi/2,\pi/2)$, we have $\left|+i\right\rangle\left|+i\right\rangle$, $\left|+i\right\rangle\left|-i\right\rangle$, $\left|-i\right\rangle\left|+i\right\rangle$, and $\left|-i\right\rangle\left|-i\right\rangle$, respectively, where $\left|{\pm}i\right\rangle=(\left|0\right\rangle\pm i\left|1\right\rangle)/\sqrt{2}$.

One can analyze the gate operation $U_{\text{CZ}}(\hat{H}\otimes\hat{H})U_{\text{CZ}}$ realized in the one-way mode [Figs.~1(c),(e)] by following the same method as that used for the target gate $(\hat{H}\otimes\hat{H})U_{\text{CZ}}$. For the box cluster state shown in Fig.~1(e), measurements on the qubits 2,3 along the basis $B_{2}^{(\alpha)}$ and $B_{3}^{(\beta)}$, respectively, will give an output state of the qubits 1,4 with
\begin{equation}
(Z\otimes X)^{s_{3}}(X\otimes Z)^{s_{2}}U_{\text{CZ}}(\hat{H}\otimes\hat{H})U_{\text{CZ}}R_{z}^{(-\alpha)}\otimes R_{z}^{(-\beta)}\left|+\right\rangle\left|+\right\rangle.\nonumber
\end{equation}
For the cases where $s_{2}=0$ and $s_{3}=0$, the post-measurement state of the 2nd and the 3rd qubits, $\left|\alpha_{+}\right\rangle_{2}\otimes\left|\beta_{+}\right\rangle_{3}$, shows the connection between input and output states of the target quantum gate by
\begin{equation}
\left|\text{Out}^{\alpha}_{\beta}\right\rangle=U_{\text{CZ}}(\hat{H}\otimes\hat{H})U_{\text{CZ}}\left|\text{In}^{\alpha}_{\beta}\right\rangle,\nonumber
\end{equation}
where $\left|\text{In}^{\alpha}_{\beta}\right\rangle=\left|-\alpha_{+}\right\rangle\left|-\beta_{+}\right\rangle$. One can design two orthonormal sets of input states which are complementary each other, for examples, the states correspond to $(\alpha, \beta)\in\{0,\pi\}$ and $(\alpha, \beta)\in\{\pi/2,-\pi/2\}$ illustrated above to analyze the gate operation.

To demonstrate the quantum gate $(\hat{H}\otimes\hat{H})U_{\text{CZ}}$ in the one-way realization, we have created states close to $\left|G'_{4}\right\rangle$ [Eq.~(2) in the main text], which is equivalent to the cluster state $\left|G_{4}\right\rangle$ up to a transformation $\hat{H}_{1}\otimes\hat{H}_{4}$. Furthermore, with the wave plate sets together with BS and PBS (Fig.~2), we perform the required measurements $B_{2}^{(\alpha)}$ and $B_{3}^{(\beta)}$ to prepare different input states of the target gate operation as describe above. See Ref.~[23] in the main text. Similarly, our created states can be directly used for implementing a quantum circuit composed of two controlled-$Z$ gates [Fig.~1(c)], since the box-cluster state [Fig.~1(e)] needed for realizing such a quantum circuit distinguishes only the state $\left|G'_{4}\right\rangle$ from a transformation $\hat{H}$ on every qubit and swap between qubits $2$ and $3$.\\

\textit{Proof of quantum steering witnesses.---}As presented in the main text, the kernel of the steering witness for a general $d$-dimensional $N$-partite and $q$-colorable graph state $\left|G\right\rangle$ is of the form
\begin{equation}
W_{N}(q,d)\equiv\sum_{m=1}^{q}P(v^{(2)}_{j}+\sum_{i\in \epsilon(j)}v^{(1)}_{i}\doteq 0|\forall j \in Y_{m}),\tag{S3}\label{wnk}
\end{equation}
for $v^{(1)}_{i},v^{(2)}_{j}\in\textbf{v}\equiv\{0,1,...,d-1\}$, where $\doteq$ denotes equality modulo $d$ and $\epsilon(j)$ represents the set of vertices that form edges with the vertex $j$ in the color set $Y_{m}$. For examples, since the state $\left|G_{4}\right\rangle$ corresponds to a chain-type two-color graph ($q=2$ and $d=2$), the corresponding witness kernel $W_{4}$ [see Eq.~(1) in the main text] then is a concrete illustration of the above representation. For the measurement setting, we design the first and second kind measurements for each party who implements quantum measurements to observables with the eigenvectors
\begin{equation}
\{\left|v^{(1)}_{k}\right\rangle_{k,1}=\left|v^{(1)}_{k}\right\rangle_{k}\},\text{ and} \ \{\left|v^{(2)}_{k}\right\rangle_{k,2}=\hat{\text{F}}_{k}^{\dagger}\left|v^{(2)}_{k}\right\rangle_{k}\}, \tag{S4}
\end{equation}
respectively. These two bases are \textit{complementary} to each other, i.e., $|_{k,1}\langle v^{(1)}_{k}|v^{(2)}_{k}\rangle_{k,2}|=1/\sqrt{d}$ for all $v^{(1)}_{k},v^{(2)}_{k}\in\textbf{v}$. Here, $\hat{\text{F}}_{k}$ is the quantum Fourier transformation defined by $\hat{\text{F}}_{k}\left|v^{(m)}_{k}\right\rangle_{k}=1/\sqrt{d}\sum_{v=0}^{d-1}\omega^{v^{(m)}_{k}v}\left|v\right\rangle_{k}$ \cite{Nielsen00}. 

The witness kernel $W_{N}(q,d)$ is designed according to the state vector of the target graph state $\left|G\right\rangle$. A truly $N$-qudit graph state \cite{Raussendorf01a,Hein04,Zhou03} can be represented by a fully-connected graph $G(V,E)$ [Fig.~1(a)]. The graph $G$ consists of the set $V$ of vertices with cardinality $|V |=N$, representing the qudits, and the set $E$ of edges each of which joins two vertices, representing interacting pairs of qudits of the graph state. A graph is called a $q$-colorable graph if the vertices of the graph $G$ can be divided into $q$ sets, say $Y_{m}$ for $m=1,2,\ldots,q$, and the vertices of each set are given a color such that adjacent vertices have different colors. An edge, $(i,j)\in E$, corresponds to an unitary two-qudit transformation among the two qudits (vertices) $i$ and $j$ by $U_{(i,j)}\!\!=\!\!\sum_{v=0}^{d-1}\left|v\right\rangle_{ii}\!\left\langle v\right|\otimes (Z_{j})^{v}$, where $\{\left|v\right\rangle_{i}\}$ is an orthonormal basis for the $i$th qudit and the operator $Z_{j}$ for the $j$th qudit is defined by $Z_{j}=\sum_{k=0}^{d-1}\omega^{k}\left|k\right\rangle_{jj}\!\left\langle k\right|$, $\omega=\exp(i2\pi/d)$. Note that $Z_{j}=Z$ for $d=2$. An explicit presentation of the state vector of the graph $G$ can be generated by applying the operators $U_{(i,j)}$ based on $G$ to an initial state $\left|f_{0}\right\rangle=\bigotimes_{k=1}^{N}\hat{\text{F}}_{k}\left|0\right\rangle_{k}$: $\left|G\right\rangle=\prod_{(i,j)\in E}U_{(i,j)}\left|f_{0}\right\rangle$. It is worth noting that all the information about the probability amplitudes of $\left|G\right\rangle$ has been considered in the steering witness kernel $W_{N}(q,d)$.

For the general preexisting-state scenario, the untrusted measurement devices of $A_{s}$ can declare random variables instead of results obtained from quantum measurements on parts of the all qudits of the genuinely multipartite entangled state. Then the extreme values of the witness kernel $W_{N}(q,d)$ is determined by the following maximization for such a situation:
\begin{equation}
W_{NC}(q,d)\equiv\max_{\mathbf{A}_{s},\{\text{v}_{a}\}_{C},\{\text{v}_{b}\}_{QM}}W_{N}(q,d),\tag{S5}
\end{equation}
where $\mathbf{A}_{s}$ denotes the index set for the subsystem $A_{s}$ for all possible bipartitions of the $N$-partite system, $\{\textbf{v}_{a}\}_{C}$ represents the set of declared random variables from $A_{s}$, and $\{\textbf{v}_{b}\}_{QM}$ is the set of outcomes derived from $B_{s}$'s quantum measurements on preexisting quantum states. Considering the concrete steering witness kernel $W_{N}(q,d)$ (\ref{wnk}) for generic $q$-colorable $N$-qudit graph states and performing the above maximization, we have
\begin{equation}
W_{NC}(q,d)=\max_{\lambda} \lambda[\hat{\text{F}}_{k}^{\dagger}\left|0\right\rangle_{kk}\!\left\langle 0\right|\hat{\text{F}}_{k}+(q-1)\left|0\right\rangle_{kk}\!\left\langle 0\right|]\nonumber
\end{equation}
\begin{equation}
\ \ \ \ =\frac{1}{2}(q+\sqrt{\frac{2(q^2-2q+2)+\gamma_{q}}{d}}),\tag{S6}\label{wnc}
\end{equation}
where $ \lambda[U]$ denotes the eigenvalues of the operator $U$, $\gamma_{2}=0$ and $\gamma_{q}=[2(q-3)+1]+\gamma_{q-1}$ for $q\geq 3$. If the experimental state $\rho_{\text{expt}}$ provides a measured witness kernel $W_{N}(q,d|\rho_{\text{expt}})$ satisfying
\begin{equation}
W_{N}(q,d|\rho_{\text{expt}})>W_{NC}(q,d),\nonumber
\end{equation}
then $\rho_{\text{expt}}$ possesses genuine $N$-partite Einstein-Podolsky-Rosen (EPR) steerability close to $\left|G\right\rangle$. Hence, we get the quantum steering witness~(6) shown in the main text.\\

\begin{figure}[t]
\includegraphics[width=8.8 cm,angle=0]{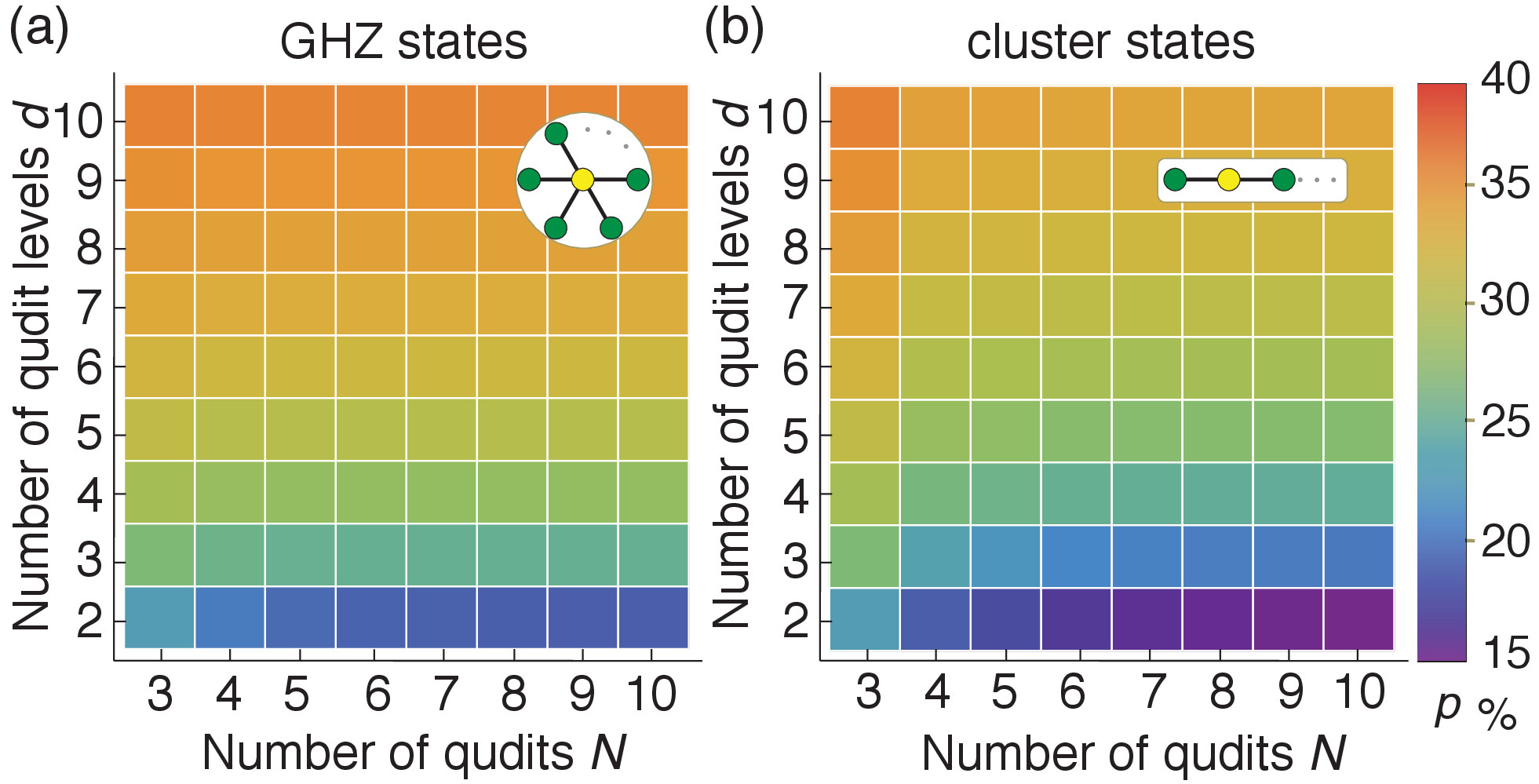}
\caption{(color online). Robustness of steering witness (6). If the probability of white nose $p_{\text{noise}}<p$ [see Eq.~(\ref{wernerstate})], then the genuine multipartite steering can be certified by the steering condition (6). Here the threshold of noise intensity $p$ is an indicator showing the noise tolerance or robustness of the steering criterion. Concrete examples of robustness of steering witness for two-color ($q=2$) star (a) and chain graph states (b) are illustrated. For large $d$, the condition (6) is robust against noise up to $p=50\%$ for both cases. }
\label{robustness}
\end{figure}  

\textit{Robustness of quantum steering witnesses.---}To investigate the robustness of our steering witness, we consider that a pure state $\left|G\right\rangle$ is mixed with white noise to be
\begin{equation}
\rho(p_{\text{noise}})=\frac{p_{\text{noise}}}{d^{N}}\hat{I}+(1-p_{\text{noise}})\left|G\right\rangle\!\!\left\langle G\right|,\tag{S7}\label{wernerstate}
\end{equation}
where $\hat{I}$ denotes the identity operator and $p_{\text{noise}}$ is the probability of uncolored noise. The mixed state $\rho(p_{\text{noise}})$ is identified to possess truly high-order EPR steerability by the steering witness~(6) if $W_{N}(q,d|\rho(p_{\text{noise}}))>W_{NC}(q,d)$. Figure~S\ref{robustness} depicts the threshold probability of uncolored noise $p$ which gives $W_{N}(q,d|\rho(p))=W_{NC}(q,d)$. We consider $p$ as an indicator of robustness of the steering witness. For large dimensionality $d$, the criterion is very robust and the noise tolerance is up to $p_{\text{noise}}<1/2$, independent of the number of qudits and the types of graph states. In addition, as illustrated by detecting states close to two-color ($q=2$) star [Greenberger-Horne-Zeilinger (GHZ) state] (a) and chain (cluster state) graph states (b), our criterion is also robust for the cases of finite $d$.\\

\textit{Experimental state fidelities $F_{S}(\rho_{\text{expt}})$.---}The steering witness (\ref{wnk}) can be represented in the operator form:
\begin{equation}
\hat{W}_{G}\equiv W_{NC}(q,d)\hat{I}-\hat{\mathcal{W}}_{N}(q,d),\tag{S8}
\end{equation}
where $\hat{\mathcal{W}}_{N}(q,d)$ is the operator presentation of the witness kernel $W_{N}(q,d)$ (\ref{wnk}) such that $W_{N}(q,d|\rho_{\text{expt}})=\text{Tr}[\hat{\mathcal{W}}_{N}(q,d)\rho_{\text{expt}}]$. If $\left\langle\hat{W}_{G}(\rho_{\text{expt}})\right\rangle=\text{Tr}[\hat{W}_{G}\rho_{\text{expt}}]<0$, then $\rho_{\text{expt}}$ is truly multipartite steerable and close to the target graph state $\left|G\right\rangle$. We use $\hat{W}_{G}$ to construct a second witness operator for detecting steerability. Let us assume that this witness operator is of the form, $\hat{W}'_{G}\equiv \delta_{G}\hat{I}-\left|G\right\rangle\left\langle G\right |$. The unknown parameter $\delta_{G}$ can be determined by showing 
\begin{equation}
\hat{W}'_{G}-\gamma\hat{W}_{G}\geq 0,\tag{S9}\label{wgcondition}
\end{equation}
where $\gamma$ is some positive constant. This relation means that when a state is detected by $\hat{W}'_{G}$, it is certified by $\hat{W}_{G}$ as well. When setting $\gamma$ as $1/2$, the above relation holds to give $\delta_{G}=W_{NC}(q,d)/2$. Hence, from the steering witness operator $\hat{W}'_{G}$, we get the criterion of genuinely $N$-partite steering on the state fidelity $F_{S}(\rho_{\text{expt}})= \text{Tr}[\left|G\right\rangle\left\langle G\right|\rho_{\text{expt}}]$
\begin{equation}
F_{S}(\rho_{\text{expt}})>\frac{1}{4}(q+\sqrt{\frac{2(q^2-2q+2)+\gamma_{q}}{d}}).\tag{S10}\label{statefidelity}
\end{equation}
For $q=2$, such as GHZ (star graph) and cluster states, the criterion reads $F_{S}(\rho_{\text{expt}})>1/2(1+1/\sqrt{d})$. Furthermore, as $d=2$ is considered for the experimental state, it provides the steering witness
\begin{equation}
F_{S}(\rho_{\text{expt}})>\frac{1}{2}(1+\frac{1}{\sqrt{2}})=\frac{1}{2}W_{4C}\sim 0.8536,\nonumber
\end{equation}
as introduced in the main text for the target state $\left|G_{4}\right\rangle$.

The steering witness in terms of $F_{S}$~(\ref{statefidelity}) is especially useful when one already has the state fidelity information obtained from experiments. One is allowed to evaluate the existing experimental results for steerability, while they did not measure the steering witness kernel before. For instances, the $N$-qubit entangled ions for $N=2,3,..,6$ created in the experiment \cite{Monz11} can be identified as genuinely $N$-partite steerable when considering their fidelities. Similarly, genuine tripartite steering can be confirmed in a superconducting circuit as well \cite{DiCarlo10}. 

With the condition~(\ref{wgcondition}), $\hat{W}'_{G}-\hat{W}_{G}/2\geq 0$, we also have
\begin{equation}
\left|G\right\rangle\left\langle G\right |\leq \frac{1}{2}\hat{\mathcal{W}}_{N}(q,d),\nonumber
\end{equation}
which implies the upper bound of the state fidelity. To derive the lower bound of the experimental state fidelity from the measured witness kernel $W_{N}(q,d|\rho_{\text{expt}})$, we use the same approach as that shown above to construct another steering witness operator, say $\hat{W}''_{G}$, which is composed of the operator $\hat{\mathcal{W}}_{N}(q,d)$. They satisfy the relation $\hat{W}''_{G}-\gamma\hat{W}'_{G}\geq 0$, i.e.,
\begin{equation}
\hat{\mathcal{W}}_{N}(q,d)-\hat{I}\leq \left|G\right\rangle\left\langle G\right |.\nonumber
\end{equation}
Hence, with the above results, the experimental state fidelity is estimated by
\begin{equation}
W_{N}(q,d|\rho_{\text{expt}})-1\leq F_{S}(\rho_{\text{expt}})\leq \frac{1}{2}W_{N}(q,d|\rho_{\text{expt}}).\tag{S11}
\end{equation}
For $d=2$, $N=4$ and $q=2$, the above result provides an useful estimation of our experimental state fidelity [see Eq.~(4) shown in the main text].  \\

\begin{figure}[t]
\includegraphics[width=7.6 cm,angle=0]{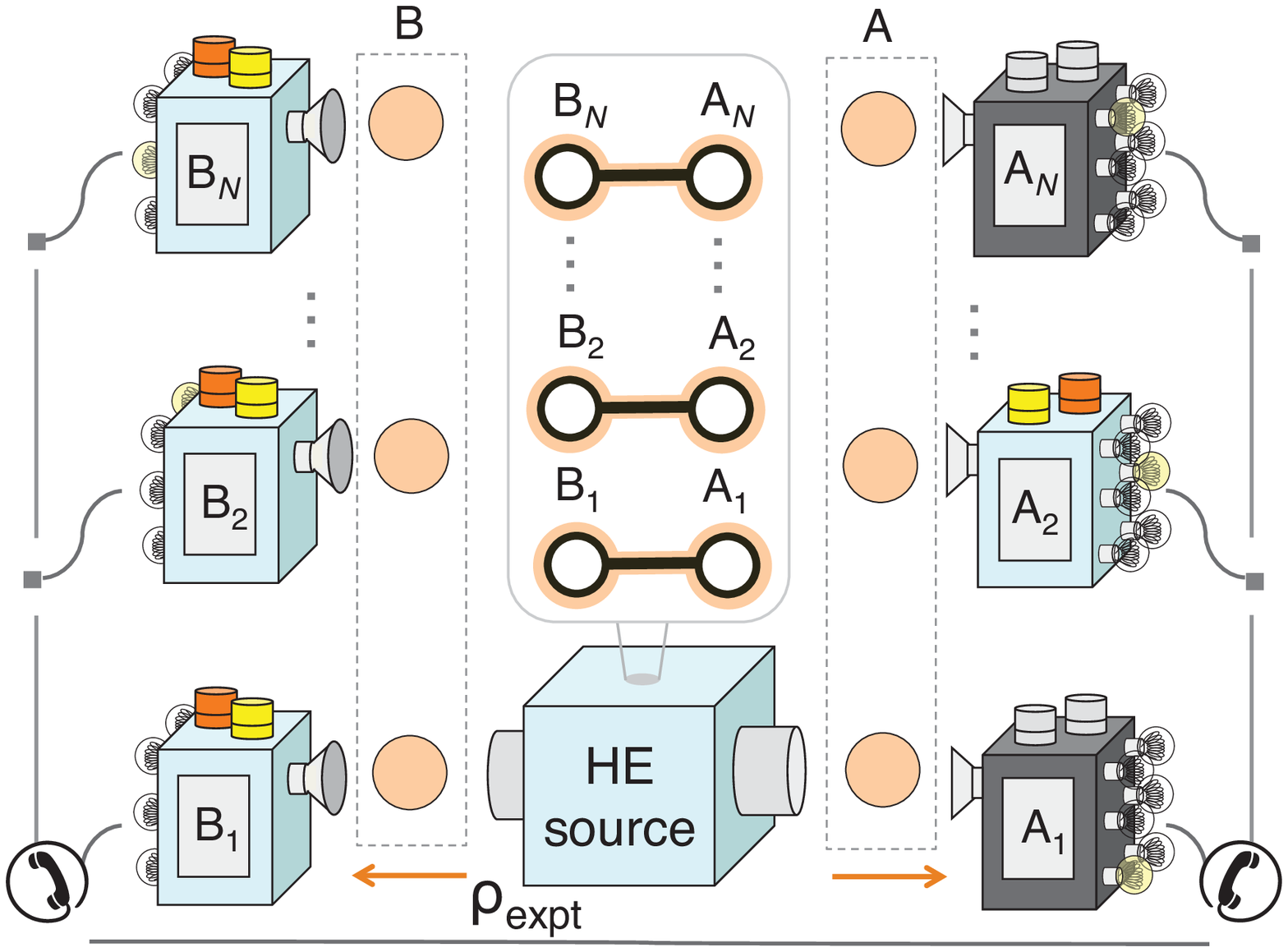}
\caption{(color online). Genuine multi-DOF EPR steering.  A hyperentanglement (HE) source \cite{Kwiat97} creates pairs of particles $A$ and $B$ which are entangled in all the $N$ DOFs, $\{A_{k},B_{k}|k=1,2,...,N\}$ \cite{Barreiro05}. We assume that the target state of such a source is $\left|H\right\rangle$ and of the form (\ref{hstate}). In practical cases, the measurement apparatuts of $A$ subsystem can be uncharacterized for some DOFs, whereas $B$ is kept by the source party under trusted measurements. When untrusted measurements are in presence of this steering scenario, the steering witness (\ref{dofsteering}) can be used to certify genuine $N$-DOF EPR steerability for states close to the target hyperentangled state $\left|H\right\rangle$. }\label{NDOF}
\end{figure}

\textit{Genuine multi-DOF EPR steering.---}We proceed to show how to observe EPR steering in \emph{all} DOFs (degrees of freedom) of a quantum system under consideration. For the main existing experimentally created $N$-DOF hyperentangled states \cite{Kwiat97,Barreiro05,Barbieri06,Vallone07}, the state in the $k$th DOF of a hyperentangled state, $\left|H\right\rangle$, with dimensionality $d_{k}$ can be described by a two-qudit graph state. One can explicitly represent such hyperentangled state as
\begin{equation}
\left|H\right\rangle=\bigotimes_{k=1}^{N}\frac{1}{d_{k}}\sum_{v=0}^{d_{k}-1}\sum_{v'=0}^{d_{k}-1}\omega^{vv'}\left|v\right\rangle_{A_{k}}\otimes\left|v'\right\rangle_{B_{k}}.\tag{S12}\label{hstate}
\end{equation}
Since this $N$-DOF state has EPR steerability in \emph{all} DOFs, the steering therein is called \emph{genuine N-DOF EPR steering}. Genuine $N$-DOF EPR steerability of a bipartite $N$-DOF system can show that the state in \emph{every} DOF is entangled even if the measurement devices for a subsystem are uncharacterized. See Fig.~S\ref{NDOF}. As illustrated, in an experiment, a hyperentanglement (HE) source \cite{Kwiat97} is aimed to create pairs of particles $A$ and $B$ which are entangled in all the $N$ DOFs, $\{A_{k},B_{k}|k=1,2,...,N\}$ \cite{Barreiro05}. The total system can be considered as $N$-pair two-qudit graph states (\ref{hstate}). In practical cases, the state preparation is imperfect. In particular, the measurement apparatuts of $A$ subsystem can be uncharacterized for some DOFs, whereas $B$ is kept by the source party under trusted measurements. Certifying the genuine multi-DOF steerability of an experimental output state $\rho_{\text{expt}}$ then becomes a critical need for such a situation.

In order to identify such new steerability, it is crucial to recognize the difference between truly $N$-DOF steering and $(N-n)$-DOF steering for $1\leq n< N$. The latter involves $n$ preexisting-state scenarios respectively for different $n$ DOFs. To rule out all of these mimicries, the following steering witness is introduced to certify genuine $N$-DOF EPR steerability for states close to the target hyperentangled state $\left|H\right\rangle$:
\begin{equation}
\prod_{k=1}^{N}\frac{1}{2}W^{(k)}_{2}(2,d_{k}|\rho_{\text{expt}})>\frac{1}{2}(1+\frac{1}{\sqrt{d}}),\tag{S13}\label{dofsteering}
\end{equation}
 where $W^{(k)}_{2}(2,d_{k})=W_{2}(2,d_{k})$ (\ref{wnk}) and $d=\min\{d_{k}\}$. This steering witness can be proven in a similar manner to the method used to prove the steering witness (6). For $n=1$, it is easy to determine the maximum value of the witness kernel as $(1+1/\sqrt{d})/2$. This happens when $1/2W^{(k')}_{2}(2,d_{k'}|\rho_{\text{expt}})=W_{2C}(2,d)=(1+1/\sqrt{d})/2$ for some DOF $k'$ with the minimum dimensionality $d_{k'}=d$ and $1/2W^{(k)}_{2}(2,d_{k}|\rho_{\text{expt}})=1$ otherwise. For any scenarios where more than one measurement devices report random outcomes ($n>1$), the maximum value of the witness kernel is smaller than the maximum value for $n=1$ because there exists more than one components with $1/2W^{(k)}_{2}(2,d_{k}|\rho_{\text{expt}})\leq W_{2C}(2,d_{k})$. Hence we prove the witness of truly $N$-DOF steering (\ref{dofsteering}).
 
In addition to the steering witness (\ref{dofsteering}), alternatively, one can utilize the information about state fidelity obtained from experiments to identify $\rho_{\text{expt}}$ as truly $N$-DOF EPR steerable. If the measured state fidelity satisfies the condition
 \begin{equation}
 F_{S}(\rho_{\text{expt}})>\frac{1}{\sqrt{d}},\tag{S14}
\end{equation}
then one can observe EPR steering in \emph{all} DOFs of $\rho_{\text{expt}}$. The method introduced to show the steering witness (\ref{statefidelity}) can be applied to the present hyperentangled systems directly.

  As reported in the pioneering experiment \cite{Barreiro05}, the experimental polarization-spatial modes hyperentangled state ($d_{1}=d_{2}=2$) has the experimental state fidelity, $F_{S}(\rho_{\text{expt}})=0.955(2)$. This satisfies the condition on the state fidelity for $d=2$, $F_{S}(\rho_{\text{expt}})>1/\sqrt{2}\sim 0.7071$, which supports the existence of genuine two-DOF EPR steerability in their experimental states. On the other hand, the proposed criteria  also can detect genuine $N$-pair EPR steering of $2N$ individual systems, such as dimer-covering states of a spin-1/2 tetramer ($N=2$). Their recent experimental demonstrations \cite{Ma11} can show such steerability by satisfying the above condition on state fidelity (see Fig.~4 in \cite{Ma11}).\\

\textit{Detecting steerability with one-way quantum computing.---}To show that genuine multipartite EPR steerability can be revealed by performing one-way quantum computing, we utilize the information about input and output of a target quantum logic gate (or quantum circuit) to construct useful steering witness (as explained by Ref. [27] in the main text).   For the quantum circuit composed of two single-qubit gates $\hat{H}$ and one two-qubit controlled-$Z$ gate, Fig.~1(b), the steering witness operator is represented by
\begin{equation}
\hat{W}_{\text{CZ}}=\delta_{\text{CZ}}\hat{I}-\hat{\mathcal{W}}_{\text{CZ}},\tag{S15}\label{wczkernel}
\end{equation}
where
\begin{equation}
\hat{\mathcal{W}}_{\text{CZ}}=\sum_{(\alpha, \beta)\in\{0,\pi\},\{\pm\frac{\pi}{2}\}}\left|\alpha_{+}\right\rangle_{22}\!\left\langle\alpha_{+}\right|\otimes\left|\beta_{+}\right\rangle_{33}\!\left\langle\beta_{+}\right| \otimes \hat{U}_{\text{CZ}}\!\left|\text{In}^{\alpha}_{\beta}\right\rangle\!\left\langle\text{In}^{\alpha}_{\beta}\right|\!\hat{U}_{\text{CZ}}^{\dag}\tag{S16}\label{wcz}
\end{equation} 
where $\hat{U}_{\text{CZ}}=(\hat{H}\otimes\hat{H})U_{\text{CZ}}$ and $\left|\text{In}^{\alpha}_{\beta}\right\rangle=\left|-\alpha_{+}\right\rangle\left|-\beta_{+}\right\rangle$ (\ref{input}). The unknown parameter $\delta_{\text{CZ}}$ is determined by following the same approach as that used to find $W_{NC}(q,d)$  (\ref{wnc}) for graph states. Here we have $\delta_{\text{CZ}}=1+1/\sqrt{2}=W_{4C}$. If the experimental results show $\langle\hat{W}_{\text{CZ}}\rangle<0$, i.e., $\langle\hat{\mathcal{W}}_{\text{CZ}}\rangle> W_{4C}$, then the state mediating such quantum gate operation is genuinely four-partite steerable. As explained by Ref. [27] in the main text and detailed in the discussions about Eqs.~(\ref{output}) and (\ref{input}), $\left|\alpha_{+}\right\rangle_{2}\otimes\left|\beta_{+}\right\rangle_{3}$ determines the input state, $\left|-\alpha_{+}\right\rangle\left|-\beta_{+}\right\rangle$, for the target gate and the connections between the input state $\left|\text{In}^{\alpha}_{\beta}\right\rangle$ and the output state $\left|\text{Out}^{\alpha}_{\beta}\right\rangle=\hat{U}_{\text{CZ}}\!\left|\text{In}^{\alpha}_{\beta}\right\rangle$. Hence, measuring the steering witness operator is equivalent to performing one-way quantum computing to analyze the eight input-output pairs of the experimental gate. Similarly, one can construct an steering witness with such feature for the box cluster state Fig.~1(c). See Eq.~(\ref{wcz2}) introduced in the next section.\\

\textit{Experimental computation fidelity $F_{\text{\text{comp}}}$.---}We proceed to show how we use $W'_{4}(\rho_{\text{expt}})$ (3) to estimate the average computation fidelity ($F_{\text{\text{comp}}}$) \cite{Hofmann05}, the quantum process fidelity ($F_{\text{process}}$) \cite{Gilchrist05}, and the average state fidelity ($F_{\text{av}}$) \cite{Gilchrist05}. Firstly, we use the operator $\hat{\mathcal{W}}_{\text{CZ}}$ as the kernel to construct a steering witness operator $\hat{W}'_{\text{CZ}}=\delta'_{\text{CZ}}\hat{I}-\hat{\mathcal{W}}_{\text{CZ}}$ such that $\hat{W}'_{\text{CZ}}-\gamma\hat{W}_{G_{4}}\geq 0$. To satisfy such relation, the unknown parameters are set as $\delta'_{\text{CZ}}=W_{4C}/2+1$ and $\gamma=1/2$. Then, by explicitly considering $\hat{W}'_{\text{CZ}}-1/2\hat{W}_{G_{4}}\geq 0$, we obtain
\begin{equation}
\hat{\mathcal{W}}_{\text{CZ}}\leq\frac{1}{2}\hat{\mathcal{W}}_{4}+\hat{I}.\tag{S17}
\end{equation}
Secondly, we construct another steering witness operator $\hat{W}'_{G_{4}}=2(W_{4C}-1)\hat{I}-\hat{\mathcal{W}}_{4}$ which satisfies $\hat{W}'_{G_{4}}-1/2\hat{W}_{\text{CZ}}\geq 0$, i.e.,
\begin{equation}
2(\hat{\mathcal{W}}_{4}-\hat{I})\leq\hat{\mathcal{W}}_{\text{CZ}}.\tag{S18}
\end{equation} 
Thus, with these results, we conclude that
\begin{equation}
2(\hat{\mathcal{W}}_{4}-\hat{I})\leq\hat{\mathcal{W}}_{\text{CZ}}\leq\frac{1}{2}\hat{\mathcal{W}}_{4}+\hat{I},\tag{S19}\label{wczassump}
\end{equation}
or, alternatively
\begin{equation}
2[W_{4}(\rho_{\text{expt}})-1]\leq\langle\hat{\mathcal{W}}_{\text{CZ}}\rangle\leq \frac{1}{2}W_{4}(\rho_{\text{expt}})+1.\tag{S20}
\end{equation}

The average computation fidelity \cite{Hofmann05} is defined by
\begin{equation}
F_{\text{comp}}\equiv\frac{1}{8}\sum_{(\alpha, \beta)\in\{0,\pi\},\{\pm\frac{\pi}{2}\}}\!\!\!\!\!\!\!\!\left\langle\text{In}^{\alpha}_{\beta}\right|\hat{U}_{\text{CZ}}^{\dag}\mathcal{E}_{\text{CZ}}(\left|\text{In}^{\alpha}_{\beta}\right\rangle\left\langle\text{In}^{\alpha}_{\beta}\right|)\hat{U}_{\text{CZ}}\left|\text{In}^{\alpha}_{\beta}\right\rangle,\tag{S21}\label{fcomp}
\end{equation}
where $\mathcal{E}_{\text{CZ}}$ denotes the experimental gate operations. Eight different input states are used to evaluate the gate performance by analyzing their respective states after the gate operation (output states). Comparing the average computation fidelity (\ref{fcomp}) with the witness kernel operator (\ref{wcz}), it is easy to find there is an important overlap between them. The average computation fidelity  $F_{\text{\text{comp}}}$ can be obtained by measuring the steering witness operator $\hat{\mathcal{W}}_{\text{CZ}}$:
\begin{equation}
F_{\text{comp}}\times \frac{1}{4}=\frac{\left\langle \hat{\mathcal{W}}_{\text{CZ}}\right\rangle}{8}.\tag{S22}\label{fcompwcz}
\end{equation}
The factor $8$ is contributed from the eight input states shown in Eq.~(\ref{fcomp}), and $1/4$ is attributed to the assumption that the ideal probability $P(\alpha_{+},\beta_{+})=1/4$ involving in Eq.~(\ref{wcz}) is assigned for all the eight settings of $(\alpha,\beta)$.

Combining the estimation (\ref{wczassump}) with the relation (\ref{fcompwcz}), as Eq.~(5) shown in the main text, one can use the experimental result $W_{4}(\rho_{\text{expt}})$ to estimate $F_{\text{comp}}$ as:
 \begin{equation}
W_{4}(\rho_{\text{expt}})-1\leq F_{\text{comp}}\leq \frac{1}{4}W_{4}(\rho_{\text{expt}})+\frac{1}{2}.\tag{S23}\label{fcompestimate}
\end{equation}
Moreover, if $W_{4}(\rho_{\text{expt}})>W_{4C}$ is experimentally ensured, we will have 
\begin{equation}
\frac{1}{2}W_{4C}< F_{\text{comp}}\leq \frac{1}{4}W_{4}(\rho_{\text{expt}})+\frac{1}{2}.\tag{S24}
\end{equation}
For the outcomes measured from our experiment, by using Eq.~(\ref{fcompestimate}), we have 
\begin{equation}
0.8829\pm 0.0049\leq F_{\text{comp}}\leq 0.9707\pm 0.0013.\tag{S25}\label{fcompestimateexp}
\end{equation}

The above results for the quantum circuit, Fig.~1(b), are applicable to the experimental gate operation $\mathcal{E}_{\text{CZ2}}$ with respect to the target quantum gate, Fig.~1(c). Similarly, one can construct a steering witness which is based on the quantum gate, Fig.~1(c), realized in the one-way mode, Fig.~1(e): $\hat{W}_{\text{CZ}2}=W_{4C}I-\hat{\mathcal{W}}_{\text{CZ}2}$ where
\begin{equation}
\hat{\mathcal{W}}_{\text{CZ}2}=\sum_{(\alpha, \beta)\in\{0,\pi\},\{\pm\frac{\pi}{2}\}}\!\!\!\!\!\!\!\!\!\!\!\!\!\!\!\!\left|\alpha_{+}\right\rangle_{22}\!\left\langle\alpha_{+}\right|\!\otimes\!\left|\beta_{+}\right\rangle_{33}\!\left\langle\beta_{+}\right|\!\otimes\! \hat{U}_{\text{CZ}2}\!\left|\text{In}^{\alpha}_{\beta}\right\rangle\!\left\langle\text{In}^{\alpha}_{\beta}\right|\!\hat{U}_{\text{CZ}2}^{\dag}\tag{S26}\label{wcz2}
\end{equation}
and $\hat{U}_{\text{CZ}2}=U_{\text{CZ}}(\hat{H}\otimes\hat{H})U_{\text{CZ}}$. Here, as used for $\hat{\mathcal{W}}_{\text{CZ}}$ (\ref{wcz}), the complementary  sets $\{\left|\text{In}^{\alpha}_{\beta}\right\rangle|\alpha,\beta=0,\pi\}$ and  $\{\left|\text{In}^{\alpha}_{\beta}\right\rangle|\alpha,\beta=\pm\pi/2\}$ \cite{Hofmann05} are utilized for the gate analysis.  Measuring the above witness operators is equivalent to performing one-way computations by inputing two sets of input states into the gates and then evaluating their results with respect to the target ones. 

To use our experimental results to estimate the average computation fidelity of this gate operation, we need a steering witness of the form Eq.~(\ref{wnk}) for the box cluster state Fig.~1(e). The kernel of this steering witness is 
\begin{equation}
W_{4Box} \equiv P(v_{1}^{(2)}+v_{2}^{(1)}+v_{4}^{(1)}\doteq 0,v_{3}^{(2)}+v_{2}^{(1)}+v_{4}^{(1)}\doteq 0)+P(v_{2}^{(2)}+v_{1}^{(1)}+v_{3}^{(1)}\doteq 0,v_{4}^{(2)}+v_{1}^{(1)}+v_{3}^{(1)}\doteq 0),\tag{S27}
\end{equation}
and the steering witness is described by
\begin{equation}
W_{4Box}(\rho_{\text{expt}})>1+\frac{1}{\sqrt{2}}.\tag{S28}
\end{equation}
Since the box-cluster state distinguishes only the experimental target state $\left|G'_{4}\right\rangle$ from a transformation $\hat{H}$ on every qubit and swap between qubits $2$ and $3$, the above witness kernel for $\left|G'_{4}\right\rangle$ then has a corresponding change in measurement settings by $W'_{4Box}\equiv P(v_{1}^{(2)}+v_{2}^{(2)}+v_{3}^{(1)}\doteq 0,v_{3}^{(1)}+v_{4}^{(1)}\doteq 0)+P(v_{2}^{(1)}+v_{3}^{(2)}+v_{4}^{(2)}\doteq 0,v_{1}^{(1)}+v_{2}^{(1)}\doteq 0)$. It is easy to find that $W'_{4Box}=W'_{4}$. Hence the value of the steering witness kernel is 
\begin{equation}
W'_{4Box}(\rho_{\text{expt}})=1.8829\pm 0.0049,\tag{S29}
\end{equation} 
which is larger than the threshold $1.7071$.

Following the same approach as shown above (\ref{wczassump}), we have 
\begin{equation}
2(\hat{\mathcal{W}}_{4Box}-\hat{I})\leq\hat{\mathcal{W}}_{\text{CZ2}}\leq\frac{1}{2}\hat{\mathcal{W}}_{4Box}+\hat{I},\tag{S30}\label{wcz2assump}
\end{equation}
i.e., 
\begin{equation}
2[W_{4Box}(\rho_{\text{expt}})-1]\leq\langle\hat{\mathcal{W}}_{\text{CZ2}}\rangle\leq \frac{1}{2}W_{4Box}(\rho_{\text{expt}})+1.\tag{S31}
\end{equation}
With this result and the same relation between $F_{\text{comp}}$ and $\langle\hat{\mathcal{W}}_{\text{CZ2}}\rangle$ as Eq.~(\ref{fcompwcz}) shown, we obtain the same lower and upper bounds of the experimental computation fidelity as described by Eq.~(\ref{fcompestimateexp}). For more extensions, the concept and method introduced here can be directly applied to generic quantum circuit of one-way quantum computing. \\

\textit{Experimental quantum process and average quantum state fidelities.---}The quantum process fidelity \cite{Gilchrist05} is a comparison between an experimental process, say $\mathcal{E}$, and a target quantum process: $F_{\text{process}}\equiv \text{Tr}[\chi_{\text{ideal}}\chi_{\text{expt}}]$, where $\chi_{\text{ideal}}$ and $\chi_{\text{expt}}$ represent the ideal and experimental process matrices \cite{Gilchrist05}, respectively. A process matrix completely characterizes the quantum process $\mathcal{E}$ \cite{Nielsen00}. While the quantum process fidelity $F_{\text{process}}$ is experimentally measurable, the number of required measurement settings increases exponentially with the number of qubits participated in a process. Fortunately, there exists a good estimation can be efficiently measured by the average computation fidelity \cite{Hofmann05}: $F_{\text{process}}\geq 2F_{\text{comp}}-1$, where two measurement settings corresponding to two complementary sets of input states are sufficient to implement the estimation. With this efficient criterion and the connection between $F_{\text{comp}}$ and $W_{4}(\rho_{\text{expt}})$ introduced above, we have 
\begin{equation}
 F_{\text{process}}\geq \frac{1}{2}W_{4}(\rho_{\text{expt}}).\tag{S32}
\end{equation}
It is also true for the quantum gate Fig.~1(c) that $F_{\text{process}}\geq W_{4Box}(\rho_{\text{expt}})/2$. Furthermore, this lower bound of process fidelity also helps derive the average quantum state fidelity \cite{Gilchrist05}, $F_{\text{av}}\equiv \text{mean}_{\text{all} \left|\psi\right\rangle}\text{Tr}\big[\mathcal{E}(\left|\psi\right\rangle\!\!\left\langle \psi\right|)\hat{U}\left|\psi\right\rangle\!\!\left\langle \psi\right| \hat{U}^{\dag}\big]$, where $\hat{U}$ is the target quantum transformation (gate operation) and the mean is over the uniform measure $\left|\psi\right\rangle$ on state space. It is know that $F_{\text{av}}$ can be represented in terms of $F_{\text{process}}$ by $F_{\text{av}}=(M F_{\text{process}}+1)/(M+1)$ \cite{Gilchrist05}, where $M$ is the dimension of the gate $\hat{U}$. Therefore we get the lower bound of average quantum state fidelity 
\begin{equation}
F_{\text{av}}\geq \frac{1}{5}[2W_{4}(\rho_{\text{expt}})+1],\tag{S33}
\end{equation}
which is applicable to the gate operation Fig.~1(c) by $F_{\text{av}}\geq \frac{1}{5}[2W_{4Box}(\rho_{\text{expt}})+1]$.\\

\textit{Quantum steering witnesses derived from full state knowledge.---}In the main text we have introduced efficient steering witnesses to certify genuine high-order steering. These steering witnesses are designed according to the features of target quantum correlations.  In what follows, we will show how to detect genuine multipartite steering of an experimental state from the full information about a target state. 

Let us assume that the target state is pure, say $\left|\psi\right\rangle$. One can design the witness kernel as:
\begin{equation}
W_{\psi}=\sum_{v_{1}^{(m_{1})}...v_{N}^{(m_{N})}}c(v_{1}^{(m_{1})}...v_{N}^{(m_{N})})P(v_{1}^{(m_{1})},...,v_{N}^{(m_{N})}),\tag{S34}
\end{equation}
where the parameters $c(v_{1}^{(m_{1})}...v_{N}^{(m_{N})})$ are derived from the tomographic decomposition of state $\left|\psi\right\rangle$ \cite{Nielsen00}: 
\begin{equation}
\left|\psi\right\rangle \left\langle\psi\right|=\sum_{v_{1}^{(m_{1})}...v_{N}^{(m_{N})}}\!\!\!\!\!\!\!\!\! c(v_{1}^{(m_{1})}...v_{N}^{(m_{N})})\left|v_{1}^{(m_{1})}\right\rangle_{1,m_{1} 1,m_{1}}\!\!\left\langle v_{1}^{(m_{1})}\right|\otimes... \otimes\left|v_{N}^{(m_{N})}\right\rangle_{N,m_{N} N,m_{N}}\!\! \left\langle v_{N}^{(m_{N})}\right|.\tag{S35}
\end{equation}
 The \emph{full} information about the target state has been used for $W_{\psi}$. Each orthonormal basis $\{\left|v_{k}^{(m_{k})}\right\rangle_{k,m_{k}}\}$ is composed of eigenvectors of a observable for the measurement $m_{k}$. Here, since the quantum state tomography is used, take $d=2$ for example, the observables can be chosen as the identity operator and three Pauli matrices. The maximum value of $W_{\psi}$ that is achieved by the preexisting-state scenario is determined by using the same method as that utilized to find $W_{NC}(q,d)$ (\ref{wnc}). We take the steering witness for the three-qubit $W$ state \cite{Dur00} as a concrete illustration of this scheme. If an experimental state $\rho_{\text{expt}}$ is detected by the following steering witness
 \begin{equation}
 W_{\psi}>\frac{1}{3}(1+\sqrt{2})\sim 0.8047,\tag{S36}
 \end{equation}
then $\rho_{\text{expt}}$ is genuinely tripartite steerable.  The above steering witness derived from the full state knowledge can be readily used to verify experimental states as truly tripartite steerable in the existing experiment, for examples, see Ref. \cite{Bourennane04}.